\pdfoutput=1
\documentclass[cernpreprint,english]{na61doc}

\usepackage{lineno}
\usepackage{colortbl}
\definecolor{darkred}{rgb}{0.5,0,0}
\definecolor{darkblue}{rgb}{0,0,0.5}
\definecolor{firebrick}{rgb}{0.75,0.125,0.125}
\definecolor{darkgreen}{rgb}{0,0.5,0}

\usepackage[colorlinks=true,linkcolor=firebrick,citecolor=darkgreen,urlcolor=darkblue]{hyperref}

\newcommand{\eV}{\ensuremath{\mbox{e\kern-0.1em V}}\xspace}
\newcommand{\GeV}{\ensuremath{\mbox{Ge\kern-0.1em V}}\xspace}
\newcommand{\MeV}{\ensuremath{\mbox{Me\kern-0.1em V}}\xspace}
\newcommand{\GeVc}{\ensuremath{\mbox{Ge\kern-0.1em V}\!/\!c}\xspace}
\newcommand{\AGeV}{\ensuremath{A\,\mbox{Ge\kern-0.1em V}}\xspace}
\newcommand{\AGeVc}{\ensuremath{A\,\mbox{Ge\kern-0.1em V}\!/\!c}\xspace}
\newcommand{\MeVc}{\ensuremath{\mbox{Me\kern-0.1em V}/c}\xspace}

\newcommand{\dd}{\ensuremath{{\textrm d}}\xspace}
\newcommand{\dedx}{\ensuremath{\dd E\!/\!\dd x}\xspace}

\newcommand{\pim}{\ensuremath{\pi^-}\xspace}
\newcommand{\pip}{\ensuremath{\pi^+}\xspace}

\newcommand{\Xim}{\ensuremath{\Xi^-}\xspace}
\newcommand{\Xip}{\ensuremath{\overline{\Xi}^+}\xspace}

\newcommand{\CernVM}{\textsc{Cern\-\kern-0.05emVM}\xspace}

\ShineTitle{Search for an exotic S=-2, Q=-2 baryon resonance in proton-proton interactions at $\sqrt{s_{NN}}$ = 17.3~\GeV}

\PreprintIdNumber{CERN-EP-20XX-XXX}

\ShineJournal{Physical Review D}
\ShineAbstract{ %
Pentaquark states have been extensively investigated theoretically in the context of the constituent quark
model. In this paper results of an experimental search for pentaquarks in the \Xim\pim, \Xim\pip, \Xip\pim and \Xip\pip invariant mass spectra in
proton-proton interactions at $\sqrt{s}$=17.3~\GeV are presented. Previous possible evidence from the NA49 collaboration of the existence of a
narrow \Xim\pim baryon resonance in p+p interactions is not confirmed with almost 10 times greater event statistics.
The search was performed using the \NASixtyOne detector which reuses the main components of the NA49 apparatus. No 
signal was observed with either the selection cuts of NA49 or newly optimised cuts.  
}
\begin{document}
\maketitle

%***********************************************************************************
\section{Introduction}
During the past decades pentaquark states have been extensively investigated theoretically  in the context of the constituent quark
model~\cite{GellMann:1964nj, Hogaasen:1978jw, Strottman:1979qu, Roiesnel:1979fj}. Some of these states are proposed to be closely bound and to have charge and strangeness quantum number combinations
that cannot be realized as three-quark states. Using the chiral soliton model an anti-decouplet of baryons was predicted
by Chemtob \cite{Chemtob:1985ar}. The lightest member was estimated by Praszalowicz~\cite{Praszalowicz:2003ik} to lie at a mass of 1530~\MeV. 
Diakonov et al.~\cite{Diakonov:1997mm} subsequently derived a width of less than 15~\MeV for this exotic baryon resonance state $\Theta^{+}(1540)$ $(uudd\bar{s})$, with S = +1, $J^P = \frac{1}{2}^{+}$. They further made predictions for the heavier members of the anti-decuplet, with the isospin quartet of
$S=-2$ baryons having a mass of about 2070~\MeV and partial decay width into $\Xi \pi$ of about 40~\MeV.
This isospin $\frac{3}{2}$ multiplet contains two $\Xi_{3/2}$ with ordinary charge assignments $\left(\Xi^0_{3/2}, \Xi^{-}_{3/2}\right)$ in addition to the exotic
states $\Xi^{+}_{3/2}$ $(uuss\bar{d})$ and $\Xi^{--}_{3/2}$ $(ddss\bar{u})$. The $\Xi_{3/2}$
isospin quartet has also been discussed as a part of higher multiplets. Jaffe and Wilczek~\cite{Jaffe:2003sg} on the other
hand based their predictions on the strong color-spin correlation force and suggest that the $\Theta^{+}(1540)$ baryon is a
bound state of two highly correlated \textit{ud} pairs and an antiquark. In their model the $\Theta^{+}(1540)$ has positive parity
and lies in an almost ideally mixed $\overline{10}_f \oplus {8}_f$ multiplet of
SU(3)$_f$ . For the isospin $\frac{3}{2}$ multiplet of $\Xi$s they predict a mass around 1750~\MeV and a width 50\% greater
than that of the $\Theta^{+}(1540)$. For the theoretical and experimental status of these low-mass multiquark states see e.g. Ref.~\cite{Liu:2014yva}. 

Experimentally, since the first observation of a $\Theta^{+}(1540)$ candidate~\cite{Nakano:2003qx}, there is still a lack of consensus about whether the lightest member of the exotic anti-decuplet has been discovered. 
After about fifteen years of excitement the results are still controversial.
There are numerous reports from different groups that conducted searches for $\Theta^{+}$ with some observing a signal and while others observing a null result, for review see Refs.~\cite{Liu:2014yva,Liu:2019zoy}. The reason why some experiments see $\Theta^{+}$, while the others do
not, may be either of experimental nature or a peculiar production mechanism (or both).

Observation of candidates for the heaviest members of the $\overline{10}$ multiplet was reported only by the NA49 experiment in p+p reactions at CERN~\cite{Alt:2003vb}. This result was not confirmed by other experiments (see e.g. Refs.~\cite{Knopfle:2004tu, Chekanov:2005at}), however, in different reactions and phase space regions. For a fuller review of the available experimental results again see Ref.~\cite{Liu:2014yva}.

An extensive program of investigation of pentaquark states containing $c$ or $b$ quarks and having masses above 4000~\MeV is being persued at many accelerators. Numerous candidates, including the ones observed recently by the LHCb experiment~\cite{Aaij:2018bre,Aaij:2019vzc}, were found and confirmed, see e.g. Ref.~\cite{Liu:2019zoy} for a recent review.

The \NASixtyOne experiment at the CERN SPS essentially reuses the detector of NA49 with upgrades allowing a factor 10 higher data recording rate. This paper discusses the experimental search for the existence of the exotic $\Xi^{--}_{3/2}$ member of the $\Xi$ multiplet employing the same detector with the same acceptance, similar analysis techniques, in the same reaction and at the same center-of-mass energy as studied by the NA49 experiment, but with 10 times greater events statistics. 
The results of the search for the $\Xi^{--}_{3/2}$ and $\Xi^{0}_{3/2}$
states and their antiparticles in proton-proton interactions at $\sqrt{s} =$17.3~\GeV are presented and compared to the published data of NA49.

%***********************************************************************************
\section{The \NASixtyOne detector}
Data used for the analysis reported here were recorded at the CERN SPS accelerator complex with the \NASixtyOne fixed target large acceptance
hadron detector~\cite{Abgrall:2014fa}, which inherited most of the apparatus from NA49. 
The \NASixtyOne tracking system consists of 4 large volume time projection chambers (TPCs). 
Two of the TPCs (VTPC1 and VTPC2) are
within superconducting dipole magnets. Downstream of the magnets two larger TPCs (MTPC-R and
MTPC-L) provide acceptance at high momenta. The interactions were produced with a beam of
158~\GeVc protons on a cylindrical liquid hydrogen target of 20~cm length and 2~cm transverse diameter.

\section{Analysis}
The recorded data sample consists of about 53M events. Reconstruction started with pattern recognition, momentum fitting, and finally formation 
of global track candidates.  These track candidates generally spanned multiple TPCs and consisted of charged particles produced in the primary interaction and at secondary
vertices.  The primary vertex was determined for each event. Events in which no primary vertex was found
were rejected. To remove non-target interactions, the reconstructed primary vertex was required to lie within the target; $\pm$9~cm in
the longitudinal ($z$) direction, and within $\pm$1~cm in the transverse
(\textit{x, y}) direction from the geometric center of the target. These cuts
reduced the data sample to 33M inelastic p+p interactions.
Particle identification was performed via measurement of the specific energy loss (\dedx) in the TPCs.
The achieved resolution is 3–6\%
depending on the reconstructed track length~\cite{Abgrall:2014fa, Aduszkiewicz:2017sei}.
The dependence of the measured \dedx on velocity was
fitted to a Bethe-Bloch type parametrisation.

%\begin{figure}[t!]
%\centering
%\includegraphics[width=.3\textwidth]{plots/$V^{0}$mass}
%\includegraphics[width=.3\textwidth]{plots/bar$V^{0}$mass}
%\caption{(Color online) The p\pim (left) and \pbar\pip (right) invariant mass spectrum
%for $V^{0}$ topologies. The gray line show the nominal $\Lambda$ masses.}
%\label{fig:LambdaMass}
%\end{figure}

The first step in the analysis was the search for $\Lambda$ candidates, which were then combined with the \pim to form
the \Xim candidates. Next, the $\Xi^{--}_{3/2}$ $\left(\Xi^{0}_{3/2}\right)$ were searched for
in the \Xim\pim (\Xim\pip) invariant mass spectrum, where the
\pim (\pip) are primary vertex tracks. An analogous procedure was followed for the antiparticles.

The $\Lambda$ candidates are formed by pairing reconstructed and identified tracks with appropriate mass assignments and opposite charge.
These particles are tracked backwards through the \NASixtyOne magnetic field from the
first recorded point, which is required to lie in one of the VTPC detectors. This backtracking is performed
in 2 cm steps in the $z$ (beam) direction. At each step the separation in the transverse coordinates $x$ and $y$ is
evaluated and a minimum is checked for. A pair is considered a $\Lambda$ candidate if the distance of closest approach
in the $x$ and $y$ direction is below 1 cm in both directions. Using the distances at the two neighbouring space
points around the point of closest approach, a more accurate $\Lambda$ position is found by interpolation. 
This position, together with the momenta of the tracks at this point, are used as the input for a 9 parameter fit using
the Levenberg-Marquardt fitting procedure~\cite{press_etal_1992}. 

%Protons and pions were selected by requiring their
%\dedx to be within $3\sigma$ around the nominal Bethe-Bloch
%value. The $\Lambda$ candidates were identified by locating the
%vertices from neutral decays (so called $V^{0}$s, mostly upstream of VTPC1). To identify these $V^{0}$s, the protons were
%paired with \pim and both tracked backwards through the
%magnetic field. The $V^{0}$ was constrained to lie on the
%trajectory of the decay particle with the largest number of VTPC clusters. 
%The $V^{0}$ position along the selected track
%and the three momentum components of both tracks
%(at that point) were found by a 4-parameter $\chi^{2}$ fit.
%% The resulting p\pim and \pbar\pip invariant mass spectrum
%%are shown in Fig~\ref{fig:LambdaMass}.

\begin{figure*}[t!]
\centering
\includegraphics[width=.3\textwidth]{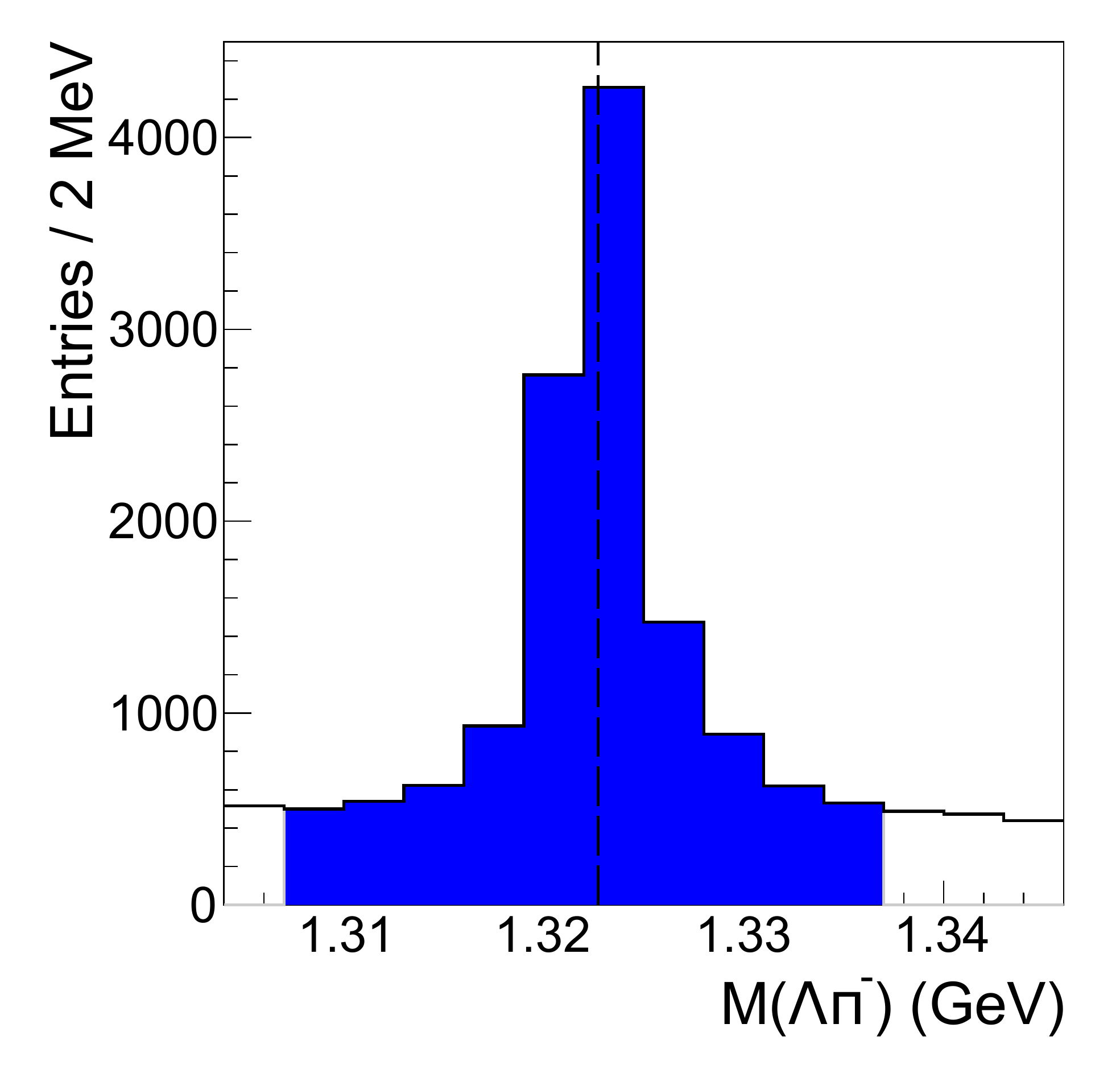}
\includegraphics[width=.3\textwidth]{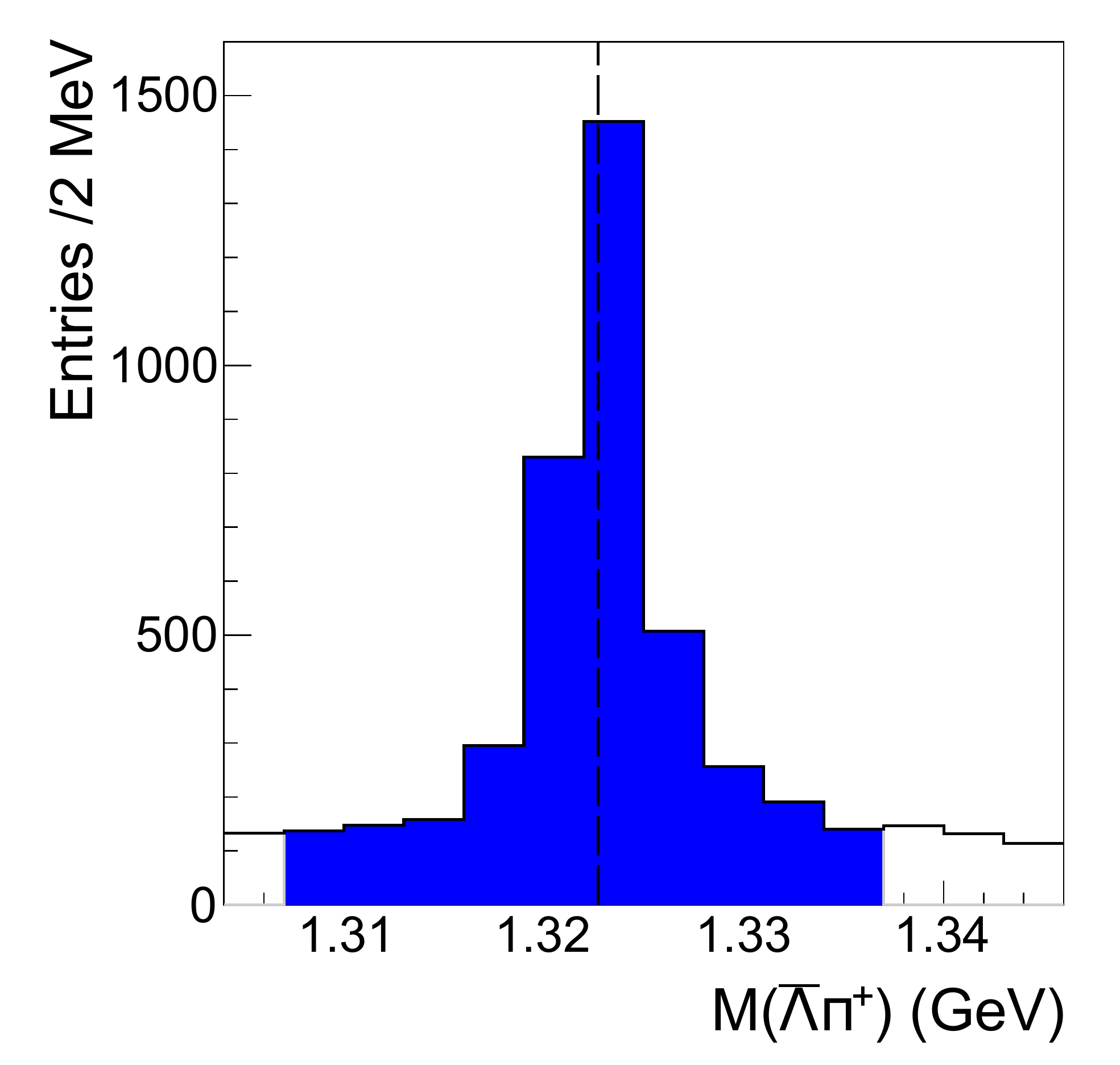}
\caption{(Color online) The $\Lambda$\pim invariant mass spectrum of
\Xim candidates (\textit{left panel}). Filled areas indicate the mass range of
the selected candidates. The vertical dashed black line shows the nominal PDG $\Xi$ mass. 
Analogous  $\bar \Lambda$\pip invariant mass spectrum of \Xip candidates (\textit{right panel}).}
\label{fig:XiMass}
\end{figure*}

\Xim candidates were assembled by the combination of all \pim with those $\Lambda$ candidates having a reconstructed
invariant mass within $\pm$15~\MeV of the nominal PDG~\cite{PhysRevD.98.030001} mass. A fitting procedure is applied using as parameters the decay position of the $V^{0}$
candidate, the momenta of both the $V^{0}$ decay tracks, the momentum of the daughter track, and finally the $z$
position of the $\Xim$ decay point. The $x$ and $y$ coordinates of the $\Xi$ decay position are not subject to the minimization,
as they are determined from the parameters using momentum conservation. This procedure yields the decay
position and the momentum of the $\Xim$ candidate. 

Specific cuts were imposed to increase the significance of the \Xim signal. As
the combinatorial background is concentrated close to the
primary vertex, a distance cut of $> 12$~cm between the
primary and the \Xim vertex was applied. Additional cuts
on extrapolated track impact parameters in the $x$ (magnetic
bending) and $y$ (non–bending) directions ($b_x$ and $b_y$) at
the primary vertex were imposed. To ensure that the \Xim
originates from the primary vertex, its $\lvert b_x \rvert$  and $\lvert b_y \rvert$ were required to be
less than 2~cm and 1~cm, respectively. On the other hand,
the \pim from the \Xim decay were required to have $\lvert b_y \rvert > 0.2$~cm.
The resulting $\Lambda$\pim invariant mass spectrum is shown in
Fig.~\ref{fig:XiMass} (\textit{left}), where the \Xim peak is clearly visible. The \Xim
candidates were selected within $\pm$15~\MeV of the
nominal \Xim mass. Only events with one \Xim candidate
(95\%) were retained. 
Exactly the same procedure was applied for antiparticles, resulting in the
\Xip peak shown in Fig.~\ref{fig:XiMass} (\textit{right}).
%The final data sample used for further analysis consisted of 13139 events containing one \Xim
%and 4114 events containing one \Xip.

\begin{figure*}[ht!]
\centering
\includegraphics[width=.49\textwidth]{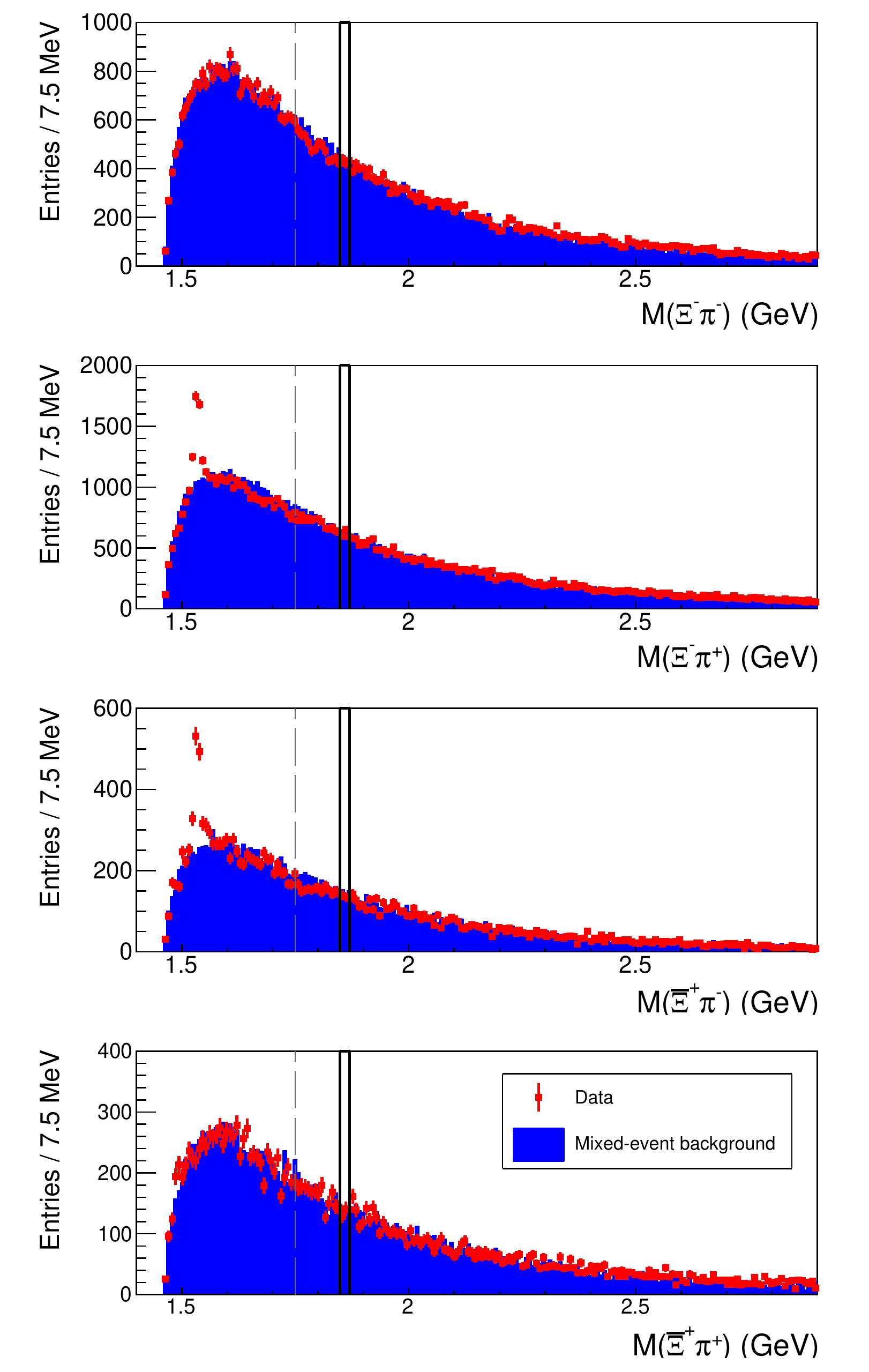}
\includegraphics[width=.49\textwidth]{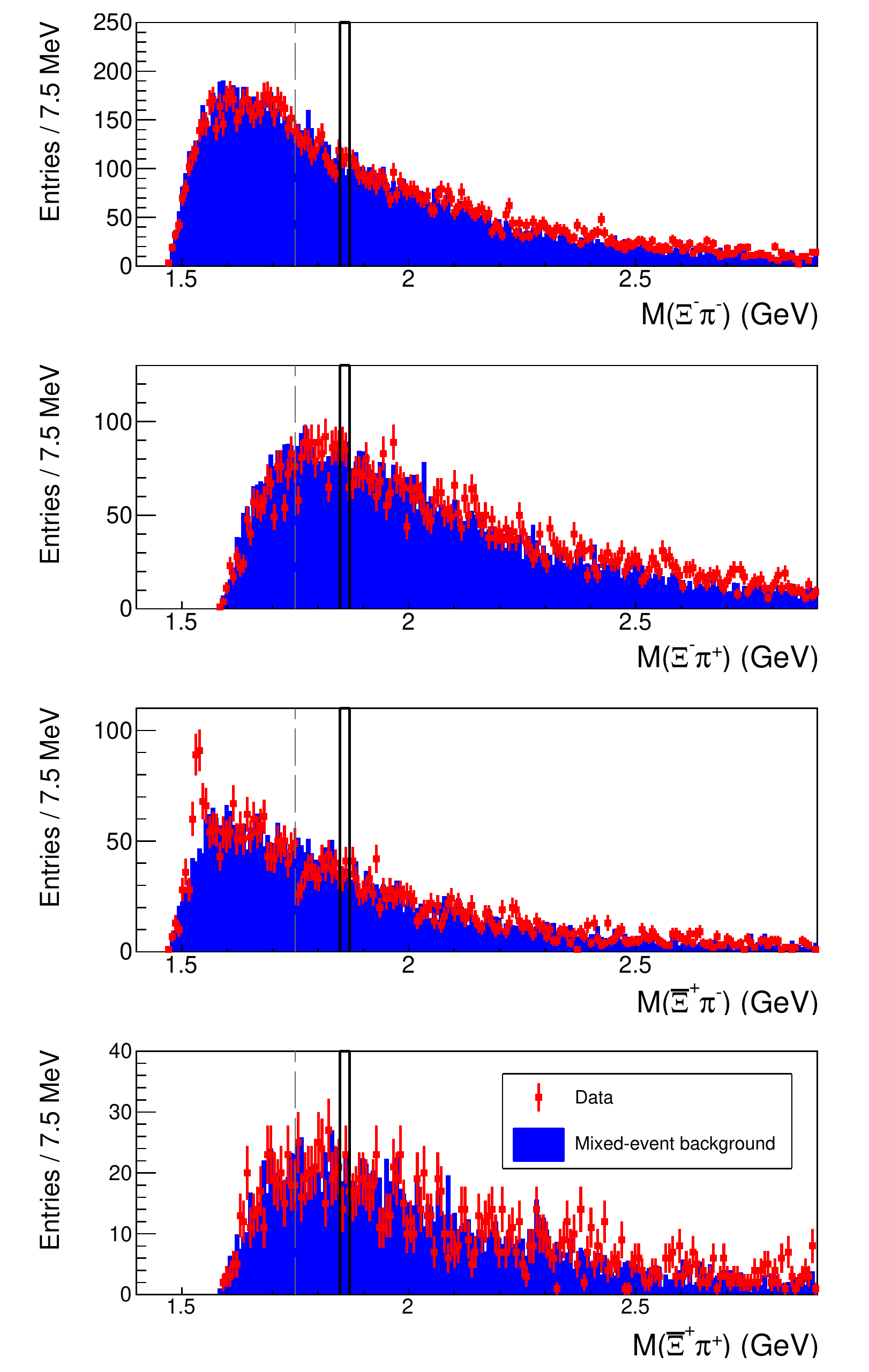}
\caption{(Color online) Invariant mass spectra of \Xim\pim, \Xim\pip, \Xip\pim, \Xip\pip combinations after selection
criteria optimized to maximize the signal to background ratio of the $\Xi(1530)$ (\textit{left}) and after selection cuts following exactly the procedure of the NA49 experiment where possible evidence of the existence of $\Xi^{--}_{3/2}$ was found (\textit{right}). The filled histograms are the normalised mixed-event background. The vertical dashed gray line shows the theoretically predicted $\Xi^{--}_{3/2}$ mass from the model of Ref.~\cite{Jaffe:2003sg}. The black rectangle indicates the mass window in which the NA49 collaboration has seen an enhancement with significance up to 4.0 standard deviations. A narrow peak of the $\Xi(1530)^0$ state is observed in the invariant mass spectra of \Xim\pip and of \Xip\pim.
}
\label{fig:penta}
\end{figure*}

To search for the exotic $\Xi^{--}_{3/2}$
the selected \Xim candidates were combined with primary \pim tracks. To select
\pim from the primary vertex, their 
%$\lvert b_x \rvert$  and 
impact parameter $\lvert b_y \rvert$ was required to
be less than 0.5~cm and their
\dedx be within 2.5$\sigma$ of their nominal Bethe-Bloch value. 
All cuts were optimized to maximize the signal-to-background ratio of the mass peaks of the $\Xi(1530)$, which decays into
the channel where the pentaquark candidates with ordinary charge assignment may be observed.
Moreover, to increase the signal-to-background ratio in the region of the $\Xi(1530)$,
an additional $\theta > 1^{o}$ cut was applied,
with $\theta$ being the opening angle between the \Xim and the \pim in the laboratory frame.
All $\Xi \pi$ combinations were analysed following the same procedure.
The resulting \Xim\pim, \Xim\pip, \Xip\pim and $\Xip\pip$ invariant mass spectra are shown
in Fig.~\ref{fig:penta}~(\textit{left}).

Additionally, a second set of more stringent selection criteria was implemented following exactly the procedure of the NA49 experiment in which possible evidence of the existence of the $\Xi^{--}_{3/2}$ was found~\cite{Alt:2003vb}:
the \Xim was required to have $\lvert b_x \rvert < 1.5$~cm  and
$\lvert b_y \rvert < 0.5$~cm at the primary vertex,
the \pim from the $\Xi^-$ decay $\lvert b_y \rvert > 0.5$~cm
at the primary vertex, and the selected \pim from the primary
vertex $\lvert b_x \rvert < 1.5$~cm  and
$\lvert b_y \rvert < 0.5$~cm. Moreover, the \dedx had to be within 
1.5$\sigma$ of the nominal Bethe-Bloch value. The  restriction on
the opening angle between the \Xim and the \pim in the
laboratory frame was $\theta > 4.5^{o}$. In addition to the described cuts, a lower cut of 3~\GeVc was imposed on the \pip
momenta to minimize proton contamination (the cut reduces the range of acceptance at small invariant mass and therefore the $\Xi(1530)$ signal disappears in the $\Xim\pip$ mass distribution). The resulting \Xim\pim, \Xim\pip, \Xip\pim, and \Xip\pip invariant mass spectra 
with NA49 selection criteria are shown in Fig.~\ref{fig:penta}~(\textit{right}).

\section{Results}

\begin{figure*}[ht!]
\centering
\includegraphics[width=.49\textwidth]{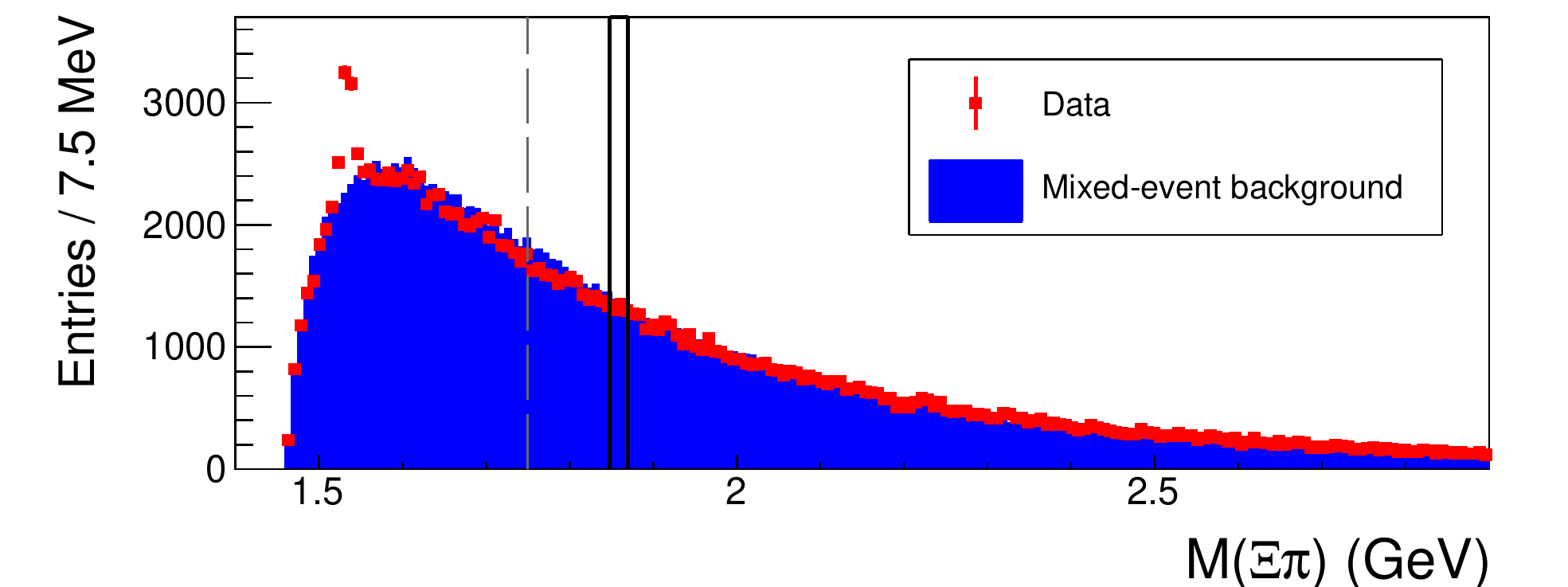}
\includegraphics[width=.49\textwidth]{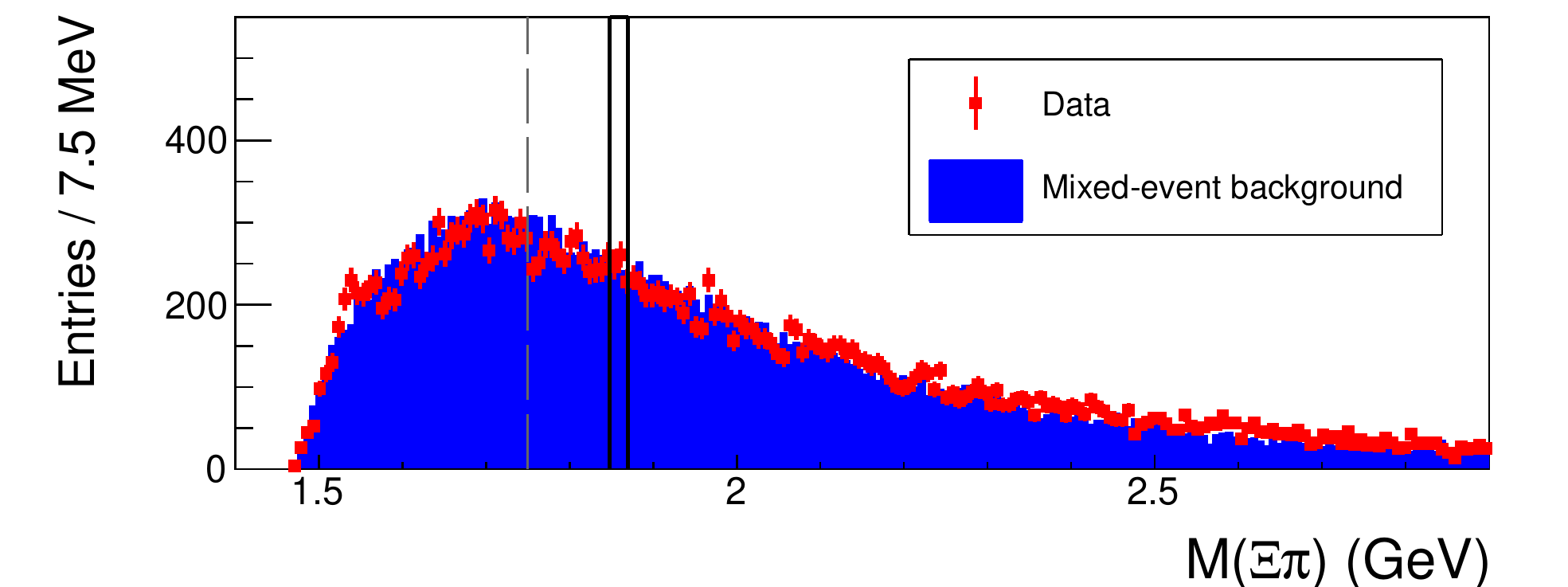}
\caption{(Color online) The sum of the \Xim\pim, \Xim\pip, \Xip\pim, and \Xip\pip invariant mass spectra  after selection
criteria optimized to maximize the signal-to-background ratio of the $\Xi(1530)$ (\textit{left}), and after selection cuts following exactly the procedure of the NA49 experiment (\textit{right}). The filled histograms are the normalised mixed-event background. The vertical dashed gray line shows the theoretically predicted $\Xi^{--}_{3/2}$ mass from the model of Ref.~\cite{Jaffe:2003sg}. The  black rectangle indicates the mass window in which the NA49 collaboration observed an enhancement with significance of 5.6 standard deviations. A narrow peak of the $\Xi(1530)^0$ state is observed.
}
\label{fig:penta_sum}
\end{figure*}

The invariant mass distributions of \Xim\pim, \Xim\pip, \Xip\pim, \Xip\pip combinations measured by \NASixtyOne are plotted in Fig.~\ref{fig:penta}. 
The filled histograms show the mixed-event background normalised to the number of real combinations. The vertical dashed gray line shows the theoretically predicted $\Xi^{--}_{3/2}$ mass from the model discussed in Ref.~\cite{Jaffe:2003sg}. The black rectangle indicates the mass window in which the NA49 collaboration has seen an enhancement with significance up to 4.0 standard deviations. For completeness, the sum of the four mass distributions is displayed in Fig.~\ref{fig:penta_sum} for both sets of cuts. For the combined 
distributions NA49 reported an observed signal significance of 5.6 standard deviations. %For comparison normalised NA49 results are presented in Fig.~\ref{fig:penta_sum} (\textit{right}).
Independently of the implemented strategy of the signal-to-background 
optimization, the data is consistent with the mixed-event background in the mass window around the theoretical predictions of the $\Xi^{--}_{3/2}$ mass. 
No signal from $\Xi^{--}_{3/2}$, $\Xi^{0}_{3/2}$ states, and their antiparticles
is observed in all invariant mass distributions.

The sensitivity of the results to variations of the different cuts and event selection criteria was investigated by varying the \dedx cut used for particle selection, by
changing the width of accepted regions around the nominal \Xim and $\Lambda$ masses, by investigating different event
topologies (e.g. the number of $\pi$ mesons per event), by selecting tracks with different number of clusters, as well as by using different $b_x$ and $b_y$ cuts. Furthermore, the influence of resonances (including the possibility of particle
misidentification) which could affect the signal
was checked. 
% by excluding them from the data. 
In all cases no signal of $\Xi^{--}_{3/2}$ emerged.

A narrow peak of the known $\Xi(1530)^0$ state~\cite{PhysRevD.98.030001} is observed in the invariant mass distribution of \Xim\pip for selection
criteria optimized to maximize the signal to background ratio of the $\Xi(1530)$, and of \Xip\pim for both selection criteria. 
The measured mass of $\Xi(1530)^0$ ($1534 \pm 3$~MeV) is consistent with PDG, while the yield scales appropriately with the number of events when comparing to NA49 results (using the NA49 selection criteria).

\begin{figure*}[ht!]
\centering
\includegraphics[width=.75\textwidth]{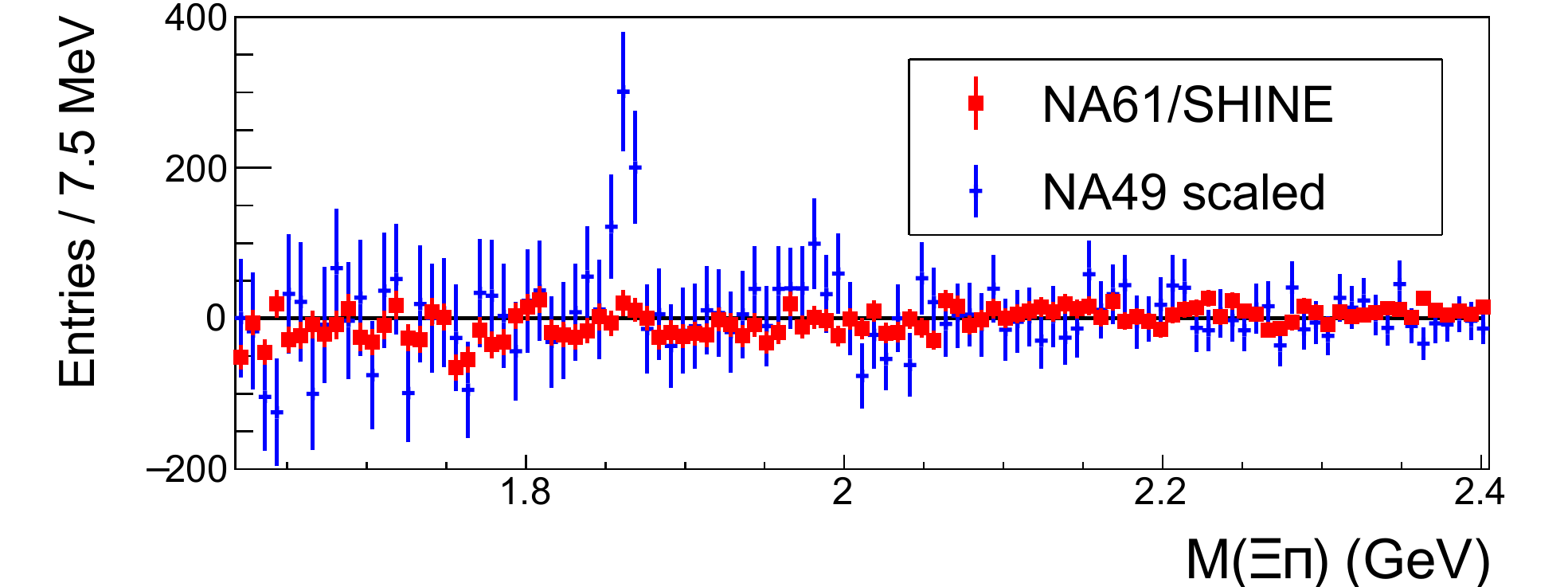}
\caption{(Color online) The background-subtracted sum of the \Xim\pim, \Xim\pip, \Xip\pim, and \Xip\pip invariant mass spectra after selection cuts following exactly the procedure of the NA49 experiment. The red squares are the \NASixtyOne data, the blue points correspond to the renormalized NA49 distribution (Fig.~3b in Ref.~\cite{Alt:2003vb}) in which a narrow peak with significance of 5.6 standard deviations is visible. 
}
\label{fig:penta_sub}
\end{figure*}

Figure~\ref{fig:penta_sub} shows the background-subtracted sum of the four invariant mass distributions displayed in Fig.~\ref{fig:penta_sum} (\textit{right}). Superimposed is a similar (Fig.~3b in Ref.~\cite{Alt:2003vb}) distribution observed by NA49 renormalized to the same number of selected p+p interactions.

In conclusion, the \NASixtyOne analysis of p+p interactions at $\sqrt{s}$=17.3~\GeV with 10 times greater statistics compared to the NA49 analysis~\cite{Alt:2003vb} does not show any indication of 
narrow $\Xi^{--}_{3/2}$,  $\Xi^{0}_{3/2}$, $\overline{\Xi}^{++}_{3/2}$,  $\overline{\Xi}^{0}_{3/2}$ states. No signal is observed in invariant mass distributions of \Xim\pim, \Xim\pip, \Xip\pim, and \Xip\pip. This is particularly true for the mass window (1848 - 1870~\MeV) in which the NA49 collaboration had seen  an enhancement with significance up to 5.6 standard deviations.%, Fig.~\ref{fig:penta_sub}.
%All four invariant mass distributions 
%shown in Fig.~\ref{fig:penta} do not show significant signals in the mass window 1.848 - 1.870~\MeV where NA49 previously found
%pentaquark candidates.
%Finally NA61 results do not show narrow peak above the background at approximately 1.86~\MeV observed by NA49. NA49 observed in the mass window 1.848 - 1.870~\MeV 81 entries
%with a background of about B = 45 events. The signal of S = 36 events had a significance of 4.0 standard deviations calculated as $S/\sqrt{S+B}$. This state was a candidate
%for the $\Xi^{--}_{3/2}$ pentaquark. Of the other 3 members of the predicted isospin quartet only the $\Xi^{--}_{3/2}$
%was observable in NA49 experiment via the \Xim\pip decay channel. Also the corresponding antibaryon states,
%$\bar{\Xi}^{++}_{3/2}$ and $\bar{\Xi}^{0}_{3/2}$ were observed.

\section*{Acknowledgments}
We would like to thank the CERN EP, BE, HSE and EN Departments for the
strong support of NA61/SHINE.

This work was supported by
the Hungarian Scientific Research Fund (grant NKFIH 123842\slash123959),
the Polish Ministry of Science
and Higher Education (grants 667\slash N-CERN\slash2010\slash0,
NN\,202\,48\,4339 and NN\,202\,23\,1837), the National Science Centre Poland (grants~2011\slash03\slash N\slash ST2\slash03691,
2013\slash11\slash N\slash ST2\slash03879, 2014\slash13\slash N\slash
ST2\slash02565, 2014\slash14\slash E\slash ST2\slash00018,
2014\slash15\slash B\slash ST2\slash02537 and
2015\slash18\slash M\slash ST2\slash00125, 2015\slash 19\slash N\slash ST2 \slash01689, 2016\slash23\slash B\slash ST2\slash00692, 
2017\slash 25\slash N\slash ST2\slash 02575, 2018\slash 30\slash A\slash ST2\slash 00226),
the Russian Science Foundation, grant 16-12-10176,
the Russian Academy of Science and the
Russian Foundation for Basic Research (grants 08-02-00018, 09-02-00664
and 12-02-91503-CERN), the Ministry of Science and
Education of the Russian Federation, grant No.\ 3.3380.2017\slash4.6,
 the National Research Nuclear
University MEPhI in the framework of the Russian Academic Excellence
Project (contract No.\ 02.a03.21.0005, 27.08.2013),
the Ministry of Education, Culture, Sports,
Science and Tech\-no\-lo\-gy, Japan, Grant-in-Aid for Sci\-en\-ti\-fic
Research (grants 18071005, 19034011, 19740162, 20740160 and 20039012),
the German Research Foundation (grant GA\,1480/2-2), the
Bulgarian Nuclear Regulatory Agency and the Joint Institute for
Nuclear Research, Dubna (bilateral contract No. 4799-1-18\slash 20),
Bulgarian National Science Fund (grant DN08/11), Ministry of Education
and Science of the Republic of Serbia (grant OI171002), Swiss
Nationalfonds Foundation (grant 200020\-117913/1), the IN2P3-CNRS (France) and the Fermi National Accelerator Laboratory (Fermilab), a U.S. Department of Energy, Office of Science, HEP User Facility. Fermilab is managed by Fermi Research Alliance, LLC (FRA), acting under Contract No. DE-AC02-07CH11359.

\bibliographystyle{na61Utphys}
%{\footnotesize\raggedright
\bibliography{na61References}
%}
\newpage
{\Large The \NASixtyOne Collaboration}
\bigskip
\begin{sloppypar}
% based on XML DB with time Tue Nov 19 12:47:47 2019
% Authors in alphabetical order.

\noindent
A.~Aduszkiewicz$^{\,15}$,
E.V.~Andronov$^{\,21}$,
T.~Anti\'ci\'c$^{\,3}$,
V.~Babkin$^{\,19}$,
M.~Baszczyk$^{\,13}$,
S.~Bhosale$^{\,10}$,
A.~Blondel$^{\,4,23}$,
M.~Bogomilov$^{\,2}$,
A.~Brandin$^{\,20}$,
A.~Bravar$^{\,23}$
% ...

\end{sloppypar}
% based on XML DB with time Tue Nov 19 12:47:47 2019
% Institutes in alphabetical order.

\noindent
% ...
$^{2}$~Faculty of Physics, University of Sofia, Sofia, Bulgaria\\
$^{3}$~Ru{\dj}er Bo\v{s}kovi\'c Institute, Zagreb, Croatia\\
$^{4}$~LPNHE, University of Paris VI and VII, Paris, France\\
% ...
$^{10}$~Institute of Nuclear Physics, Polish Academy of Sciences, Cracow, Poland\\
% ...
$^{13}$~AGH - University of Science and Technology, Cracow, Poland\\
% ...
$^{15}$~University of Warsaw, Warsaw, Poland\\
% ...
$^{19}$~Joint Institute for Nuclear Research, Dubna, Russia\\
$^{20}$~National Research Nuclear University (Moscow Engineering Physics Institute), Moscow, Russia\\
$^{21}$~St. Petersburg State University, St. Petersburg, Russia\\
% ...
$^{23}$~University of Geneva, Geneva, Switzerland\\
% ...

\end{document}